\documentstyle[11pt,epsf]{article}

\newcommand{\beq}{\begin{equation}}
\newcommand{\eeq}{\end{equation}}
\newcommand{\bea}{\begin{eqnarray}}
\newcommand{\eea}{\end{eqnarray}}
\newcommand{\is}{ &\!=\!& }
\newcommand{\isim}{ &\!\sim\!& }

\newcommand{\Gt}{{\widetilde{G}}}
\newcommand{\Dt}{{\widetilde{D}}}
\newcommand{\cdir}{c^{{\rm Dir}}}
\newcommand{\csi}{c^{{\rm SI}}}

\renewcommand{\theequation}{\thesection.\arabic{equation}}

\newcounter{subequation}[equation]
\makeatletter
\expandafter\let\expandafter\reset@font\csname reset@font\endcsname
 
\def\subeqnarray{\arraycolsep1pt
    \def\@eqnnum\stepcounter##1{\stepcounter{subequation}
        {\reset@font\rm(\theequation\alph{subequation})}}\eqnarray}

\makeatother
 
\begin{document}
\thispagestyle{empty}
\parskip=12pt
\raggedbottom

\renewcommand{\thefootnote}{\fnsymbol{footnote}}
\def\mytoday#1{{ } \ifcase\month \or
 January\or February\or March\or April\or May\or June\or
 July\or August\or September\or October\or November\or December\fi
 \space \number\year}
\noindent
\hspace*{9cm} MPI--PhT/96--121\\
\hspace*{9cm} BUTP/96--27\\
\vspace*{1cm}
\begin{center}
{\LARGE Questionable and unquestionable in the \\[3mm]
perturbation theory of non-Abelian models}%
\footnote{Work supported in part by Schweizerischer Nationalfonds}

\vspace{1cm}

Ferenc Niedermayer\footnote{On leave from the Institute of Theoretical
Physics, E\"otv\"os University, Budapest, Hungary}
\\
\vskip 1ex
Institute for Theoretical Physics \\
University of Bern \\
Sidlerstrasse 5, CH-3012 Bern, Switzerland

\vspace{0.5cm}

Max Niedermaier and Peter Weisz\\
\vskip 1ex
Max-Planck-Institut f\"ur Physik\\
F\"ohringer Ring 6, D-80805 M\"unchen, Germany \\

\mytoday \\ \vspace*{0.5cm}

\nopagebreak[4]

\begin{abstract}
\noindent We show, by explicit computation, that bare lattice 
perturbation theory in the two-dimensional ${\rm O}(n)$ nonlinear 
$\sigma$ models with superinstanton boundary conditions is divergent 
in the limit of an infinite number of points $|\Lambda|$. 
This is the analogue of David's statement that renormalized 
perturbation theory of these models is infrared divergent in
the limit where the physical size of the box tends to infinity.
We also give arguments which support the validity of the bare 
perturbative expansion of short-distance quantities obtained by 
taking the limit $|\Lambda|\to\infty$ term by term in the theory 
with more conventional boundary conditions such as Dirichlet, 
periodic, and free.
\end{abstract}

\end{center}
\eject

\renewcommand{\thefootnote}{\arabic{footnote}}
\setcounter{footnote}{0}

\section{Introduction}

Quantum chromodynamics (QCD) is the presently favored candidate theory
for strong interactions. However, to establish its esteemed status
the theory must be able to reproduce the low-lying spectrum and 
low-energy $S$-matrix elements to a respectable precision. 
To accomplish this we require a nonperturbative definition 
of the theory -- the most promising of which is lattice 
regularization. On the other hand, many aspects of high energy 
phenomena involving hadrons (e.g., jets, deep inelastic scattering, etc.) 
are described successfully using renormalized perturbation theory (PT); 
the rationale being the expected property of asymptotic freedom, 
i.e., that the amplitudes can be expressed as a series in a coupling 
which depends on the energy of the process and goes to zero as the 
energy goes to infinity. The conventional wisdom is that 
``in principle" one could justify this use of PT by showing that 
the coefficients of the conventional perturbative series provide 
the true asymptotic expansion of the full theory defined 
nonperturbatively via the lattice regularization. Confirming this 
proclaimed status of PT is also essential in aiming at a criterion 
where to truncate it. Without such a criterion PT produces sequences 
of predictions, one for each order where one decides to truncate it, 
and {\it a priori} none of them might be close to the true answer.
It must be emphasized, however, that the usual arguments for the above
scenario are far off a proof and  Patrascioiu and Seiler 
repeatedly emphasized that also alternative scenarios can be 
imagined \cite{PS,PSI}.

In full QCD a rigorous understanding of the status of PT is likely not 
to become amenable in the near future, so that it is sensible to address 
the question in a simpler context, e.g., the O($n$) nonlinear $\sigma$ models
in two dimensions. These models are also perturbatively asymptotically free
and are thought to describe a multiplet of stable massive particles,
the mass scale being nonperturbative in the coupling constant.
In the lattice formulation one usually starts with a finite number of 
lattice points $|\Lambda|$ and the standard discretization (\ref{1}). 
The conventional picture is that the critical point at which the 
continuum limit for $n>2$ should be taken is $\beta_c=\infty$. 
Patrascioiu and Seiler \cite{PS}, on the other hand, conjecture that 
there is a critical $\beta_c<\infty$ (for all $n>1$) beyond which 
the theory is in a massless phase. It might then happen that although 
the continuum limit obtained by approaching $\beta_c$ from below has 
a mass gap and describes the expected low-energy properties, the 
theory thus obtained is not asymptotically free at high energies. In 
particular, conventional perturbation theory in the infinite volume 
limit would then be wrong for 2-d models with a non-Abelian global 
symmetry.
Their arguments are based on results from percolation theory \cite{PSI}
and with analogy to the case of the Abelian model $n=2$ (for which it 
is rigorously proven that there is a phase transition at finite $\beta$ 
\cite{Fro}). Numerical investigations can give useful hints but 
obviously cannot establish which scenario is correct.

Concerning the status of perturbation theory it is known that for a 
fixed number of lattice points $|\Lambda| \sim L^2$ and for a large 
class of boundary conditions (BC's) observables have a well-defined 
perturbative expansion in the bare coupling $\beta^{-1}$. 
We shall use the term ``short-distance observable'' to denote 
O($n$)--invariant correlation functions of some local field, where 
(in the bulk of the paper) all arguments are fixed lattice sites, a
distance $O(L)$ away from the boundary. The problems now arise with the 
limit $L\to \infty$. Let us first consider the following question:
\begin{itemize}
\item[{\bf Q1:}] Is the asymptotic expansion of short-distance 
observables for fixed $L$ and $\beta\to\infty$ uniform in $L$? 
Equivalently, can for such quantities the limits $\beta \to \infty$ and 
$L\to \infty$ be exchanged in bare lattice PT?
\end{itemize}
A major finding of \cite{PSII} is that the answer to {\bf Q1} 
actually depends on the choice of boundary conditions.
The reasoning can be illustrated with the computation of the energy 
density ${\cal E}$ at the center of the lattice. 
Although ${\cal E}$ does not have a continuum limit relevant
to the physics, by application of the Mermin-Wagner theorem
\cite{MW} it should be independent of the boundary conditions 
in the limit of an infinite number of lattice points. 
Patrascioiu and Seiler invented boundary conditions (which they 
coined superinstanton (SI) boundary conditions; c.f.~sect.~2) 
and computed the one-loop coefficient of ${\cal E}$. 
They showed that it has a finite limit as 
$L\to\infty$ but the result was different from that with more 
conventional BC's such as Dirichlet. 
Furthermore they claim a similar result holds for the renormalization 
group $\beta$-functions. The conclusion is that for at least 
{\it one} of the BC's involved the answer to {\bf Q1} is negative. 
The authors interpret their result as casting doubt on the validity of 
standard PT for {\it all}\, BC's. They claim that
the correct perturbative treatment in the limit of infinite volume
must include expansions around the so-called superinstanton gas
(configurations whose energy tends to zero as
$L\to\infty$; see sect.~3).

We believe that such conclusions are too strong and we set
out to clarify some of the points raised. In the next section we show
that the SI BC are ``sick" in the sense that the limits in {\bf Q1} 
cannot be interchanged. We do this by showing that 
the perturbative coefficients of ${\cal E}$ with SI BC diverge 
at two-loop order in the limit $L\to\infty$.%
\footnote{We have verified, however, that
if we compute quantities well away from the boundaries and the center
then the same results are obtained as for conventional BC.}
We have not considered the particular
$\beta$-functions  defined in \cite{PSII} but we believe a similar
result will be found.
This infrared divergence of the PT with SI BC was already
suspected by David \cite{DavidComment} but as his argument was
formulated in the framework of (renormalized) continuum PT his claims 
were open to a counterattack \cite{PSreply} and the issue 
remained unsettled. Our result can be viewed as the lattice analogue 
of David's statement.

Having identified the SI BC as ``sick", the next question 
is whether the conventional BC are ``healthy". Consider the bare 
lattice PT expansion of some short-distance observable on a 
finite lattice $|\Lambda| \sim L^2$ with various BC's. Let
$c^{\alpha}_r(x_1,\ldots, x_n;L),\;r\geq 1$ be the coefficient
of $\beta^{-r}$ in an asymptotic expansion with BC of type 
$\alpha$, where the sites $x_1,\ldots,x_n$ are fixed and a 
distance O($L$) away from the boundary.  
\begin{itemize}
\item[{\bf Q2:}] Which BC's $\alpha$ can be anticipated to give
finite and coinciding answers for the $L\to\infty$ limits of their
perturbative coefficients $c^{\alpha}_r(x_1,\ldots, x_n;L)$?
\item[{\bf Q3:}] Do the infinite volume short-distance observables 
admit an asymptotic expansion in $\beta^{-1}$? If ``yes", can their 
coefficients $c_r^{\infty}(x_1,\ldots,x_n),\;r\geq 1$ be obtained
from PT via $\lim_{L\to \infty} c^{\alpha}_r(x_1,\ldots,x_n;L)=
c_r^{\infty}(x_1,\ldots,x_n)$, where $\alpha$ is one of the BC
meeting the conditions in {\bf Q2}? 
\end{itemize}
In section 3 we shall consider these deeper questions raised in 
\cite{PS,PSII} and argue in favor of the following picture: 
(i)~All BC's involving only ferromagnetic couplings and leaving 
the interior spins unconstrained meet the condition in {\bf Q2}
-- provided free and Dirichlet BC meet it.
(ii)~Assuming the latter also the answer to both parts of {\bf Q3} 
is affirmative. We do not have a rigorous theorem but our considerations
may constitute a strategy for a future proof thereof. 
Section 4 is devoted to some further discussion.

\section{IR divergence of PT with superinstanton BC}

We consider a two-dimensional square lattice
$\Lambda=\bigl\{(x_1,x_2);-L/2+1\le x_k\le L/2-1\bigr\}$, $L$
a positive even integer. The set of points surrounding the square, 
$\bigl\{ |x_1|=L/2,~ |x_2| \le L/2 \bigr\}  \cup
 \bigl\{ |x_1| \le L/2,~ |x_2| = L/2 \bigr\}$ will be referred to 
as the boundary $\partial\Lambda$ of $\Lambda$. The
${\rm O}(n)$ spin field ${\bf S}_x$ defined on $\Lambda \cup \partial
\Lambda$ is an $n$-component field of unit length ${\bf S}_x^2=1$.
We restrict attention to the standard lattice action
\beq
{\cal A}(S)=\beta \sum_{x,\mu} 
\left( 1-{\bf S}_x \cdot {\bf S}_{x+\mu} \right).
\label{1}
\eeq
For fixed $L$ the perturbative expansion of the two-point function is
\beq
C^{\alpha}(x,y;\beta,L)\equiv
\langle {\bf S}_x \cdot {\bf S}_y \rangle = 
1 - \sum_{r\geq 1} \beta^{-r} c_r^{\alpha}(x,y;L)\;,
\eeq
where the superscript ${\alpha}$ indicates the dependence on the 
boundary conditions under consideration. In this section we will only 
consider two sorts of BC's, the Dirichlet boundary condition 
$S_x^a=\delta_{an}$ for $x\in\partial\Lambda$ and the so-called 
superinstanton (SI) boundary conditions \cite{PSII}.
The latter are Dirichlet BC's with the additional constraint that the
field at some point $z_0$ is also held fixed parallel to the fields 
at the boundary.

The purpose of this section is to show, by explicit computation, that
PT with SI BC's is infrared (IR) divergent at third order, while
PT with Dirichlet or periodic BC is IR finite at the same order.  
For completeness, and to introduce the notations, we first reconsider
the first and second orders in some detail.
Bare perturbation theory can be performed by parametrizing the
fields by%
\footnote{This is not indispensable, a parametrization independent 
definition could be given via the Schwinger-Dyson equations.}
$S_x^a=\beta^{-\frac12}\pi_x^a,\,a=1,\dots,N$ where
$N=n-1$, and $S_x^n=(1-\beta^{-1}\vec{\pi}_x^2)^{\frac12}$.
The measure $\Pi_{x\in\Lambda}d{\bf S}_x\delta({\bf S}_x^2-1)$
is then replaced by 
$\exp(-{\cal A}_{\rm measure})\Pi_{x\in\Lambda}d\vec{\pi}_x$ with
${\cal A}_{\rm measure}=\frac12\sum_x\ln(1-\beta^{-1}\vec{\pi}_x^2)$.
The Feynman diagrams contributing to the two-point function are the same
for the two BC's, only the expression for the free propagator differs.
We denote the free Dirichlet propagator by $D(x,y)$ and the free
SI propagator by $G(x,y)$. One has
\beq
G(x,y) = D(x,y) - \frac{D(x,z_0) D(z_0,y)}{D(z_0,z_0)}\;.
\label{Gxy}
\eeq
In the following we will take the point $z_0$ to be the origin
$z_0=(0,0)$. Also we denote by $z_1$ the point $(1,0)$.
One easily sees that each $c_r^{\alpha}(x,y;L)$
is a polynomial in $N=n-1$
of degree $r$ or less, and we write
\beq
c^{\alpha}_r=\sum_{j=1}^r c^{\alpha}_{r;j}N^j\,.
\eeq

\subsection{First order}

In first order we have, for the Dirichlet BC,
\beq
\cdir_1(x,y)=\frac12 N \Bigl[ D(x,x)+D(y,y)-2D(x,y) \Bigr]\;.
\eeq
As mentioned above, for the SI BC one simply replaces $D$ by $G$.
For the special case of the energy expectation value
we have
\beq
\cdir_1(z_0,z_1)=N\Bigl(\frac14-\frac12 \triangle_0\Bigr) \;,
\label{c1Dir}
\eeq
where
$$
\triangle_0=D(z_0,z_0)-D(z_1,z_1).
$$
$\cdir_1$ approaches its asymptotic value with power corrections since
$\triangle_0=O(1/L^2)$ for $L\to\infty$. On the other hand
\beq
\csi_1(z_0,z_1)=N\frac12 G(z_1,z_1)
\eeq
approaches the same asymptotic value extremely slowly with corrections
$1/\ln L$ since
\beq
G(z_1,z_1)=\frac12-\triangle_0-\frac{1}{16}Z^{-1}\;,
\label{Gz1z1}
\eeq
with
\beq
Z\equiv D(z_0,z_0)\sim (2\pi)^{-1} \ln L \;, \mbox{~~~for~}
      L\to\infty \;.
\eeq

\subsection{Second order}

\begin{figure}[htb]
\begin{center}
\leavevmode
\epsfxsize=90mm
\epsfbox{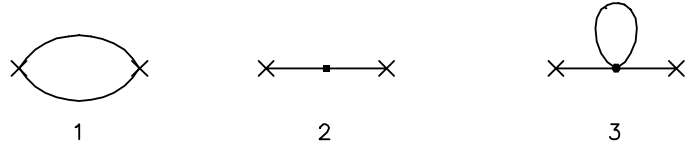}
\vskip 10mm
\end{center}
\caption{One-loop diagrams contributing to the spin-spin correlation
 function}
\label{fig:1loop}
\end{figure}

At the next order there are three contributions
\beq
c_2^{\alpha}(x,y)=\sum_{r=1}^3 c_2^{\alpha}{~}^{(r)}(x,y)
\eeq
corresponding to the diagrams depicted
in Fig.~1 (diagram 2 comes from the measure part of the action).
One has
\begin{subeqnarray}
\cdir_2{~}^{(1)}(x,y) & = &\phantom{-} 
\frac18 N^2\Bigl[ D(x,x)-D(y,y)\Bigr]^2 \nonumber\\
&&
+\frac14 N\Bigl[ D(x,x)^2+D(y,y)^2-2D(x,y)^2\Bigr]\;,
\\
\cdir_2{~}^{(2)}(x,y) & = & \phantom{-}\frac12 N\sum_i \Dt(x,y;i)^2\;,
\\
\cdir_2{~}^{(3)}(x,y) & = &  -\frac14 N^2\sum_{\langle i,j \rangle}\;
[D(i,i) -D(j,j)][\Dt(x,y;i)^2-\Dt(x,y;j)^2]
\nonumber \\ 
&& -\frac12 N \sum_{\langle i,j \rangle}\; \Bigr\{\Dt(x,y;i)
[D(i,i)\Dt(x,y;i)-D(i,j)\Dt(x,y;j)]
\nonumber \\ 
&&\makebox[2.1cm]{ }+(i\leftrightarrow j)\Bigr\}\;,
\end{subeqnarray}
where the sums in the last equation are restricted to the case when
$i$ and $j$ are nearest neighbors and we have introduced the notation
$$
\Dt(x,y;i) = D(x,i)-D(y,i).
$$
Let us just inspect the coefficient of $N^2$. Here diagram 2 does not
contribute and for diagram 1 one has simply
\beq
\csi_{2;2}{~}^{(1)}(z_0,z_1)-\cdir_{2;2}{~}^{(1)}(z_0,z_1)
=\frac18[G(z_1,z_1)^2-\triangle_0^2]
\eeq
which by eq.~(\ref{Gz1z1}) tends to $1/32$ as $L\to\infty$.
For diagram 3 we find
\beq
\csi_{2;2}{~}^{(3)}(z_0,z_1)-\cdir_{2;2}{~}^{(3)}(z_0,z_1)
= \frac{1}{16}\triangle_0
+t_1Z^{-1}+t_2Z^{-2}+t_3Z^{-3}\;,
\label{c22diff}
\eeq
with
\bea
t_1\is \phantom{-}\frac14\sum_{i\ne z_0} \Dt(z_0,z_1;i)^2E(i)\;,
\nonumber\\
t_2\is -\frac{1}{64}\sum_{i\ne z_0} D(z_0,i)^2F(i)\;,
\nonumber\\
t_3\is \phantom{-}\frac{1}{64}\sum_{i\ne z_0} D(z_0,i)^2E(i)\;,
\eea
where
\bea
E(i)\is\sum_{j=\langle i\rangle}
\Bigl[ D(i,z_0)^2-D(j,z_0)^2 \Bigr]\;,
\nonumber \\
F(i)\is\sum_{j=\langle i\rangle}\Bigl[ D(i,i)-D(j,j) \Bigr]\;.
\eea
Here the sums are taken over sites $j$  which are nearest neighbors 
to $i$. The decomposition in the form of eq.~(\ref{c22diff}) is 
made in order to be able to extract the asymptotic behavior reliably.
Our ansatz (which we have not proven analytically) is that
the functions $t_i$ have an expansion of the form
\beq
t_i(L)\sim\sum_{r=0}^{R_i}
t_i^{(r)} [(2\pi)^{-1}\ln L]^r +O(1/L),
\eeq
with $R_i$ finite. To obtain the leading behavior we have computed 
$t_i(L)$ over a large range of $L$ (up to $L=220$, using extended 
precision arithmetics) and taken logarithmic derivatives with respect 
to $L$. Our findings are that
\beq
R_1=R_2=0\;,
\eeq
and hence the contributions in eq.~(\ref{c22diff}) involving 
$t_1,t_2$ vanish in the limit $L\to\infty$. However,
\beq
R_3=3,\;\;\; t_3^{(3)}=-{1\over 96}\;.
\eeq
One can in fact understand the result for $t_3$ analytically in
the following way. Far from the origin and the boundary, i.e., for 
$1 \ll |i|\ll L$  the Dirichlet propagator is
well approximated by its continuum behavior
\beq
D(z_0,i)\sim (2\pi)^{-1}\ln {L\over |i|},
\label{Dz0i}
\eeq
and so
\beq
E(i)\sim -\Box_{(i)}(D(z_0,i)^2)\sim -2(2\pi)^{-2}|i|^{-2},
\label{Ei}
\eeq
where $\Box_{(i)}$ is the laplacian in the variable $i$.
Thus for $t_3$ we have
\bea
t_3\isim -{1\over 32} (2\pi)^{-4} {1\over L^2}\sum_{i \neq z_0}
\Bigl({L\over |i|}\Bigr)^2\ln^2{L\over |i|}
\nonumber \\
\isim -\frac{1}{32} (2\pi)^{-4}
\int_{|x|>1/L}^{|x|=1} {\rm d}^2x |x|^{-2}
\ln^2|x| = -\frac{1}{96}(2\pi)^{-3}\ln^3 L \;.
\eea
This is the same result as was guessed in \cite{PSII}.
These authors were quite fortunate to get the correct 
result because initially they just evaluated the full function
$\csi_{2;2}{~}^{(3)}$ numerically and then made a rather naive 
extrapolation. We stress that the analysis has to be done extremely 
carefully on the lines outlined above to properly treat the appearance 
of polynomials of logarithms in the denominators.%
\footnote
{Indeed, $\csi_{2;2}{~}^{(3)}$ is not monotonic in $L$ and has a minimum
at a large value of $L\sim 120$.}
 
The final result of this subsection is that
\beq
\lim_{L\to\infty}[\csi_{2;2}(z_0,z_1)-\cdir_{2;2}(z_0,z_1)]={1\over 48}
\eeq
(we also know $\lim_{L\to\infty}\cdir_{2;2}(z_0,z_1)=0$).
The Mermin-Wagner theorem implies that the perturbative expansion of 
the two-point function is independent of the boundary conditions in
the infinite volume limit. Accepting this fact we must conclude that 
the interchange of the limits $\beta\to\infty$ and $L\to\infty$ is not 
permissible for at least one of the boundary conditions involved. 
We now go on to show that this is indeed the case for the SI BC
because one encounters an infrared divergence.

\subsection{Leading $N$ contribution to the third order}

For the purpose of showing an IR divergence at some
order $r$, for generic $n$,
it is sufficient to show that one of the coefficients of this
polynomial in $N$ is IR divergent. 
We claim that this is the case for the
coefficient of $N^3$ in $c_3^{\alpha}(x,y;L)$ with the SI BC.

\begin{figure}[htb]
\begin{center}
\leavevmode
\epsfxsize=90mm
\epsfbox{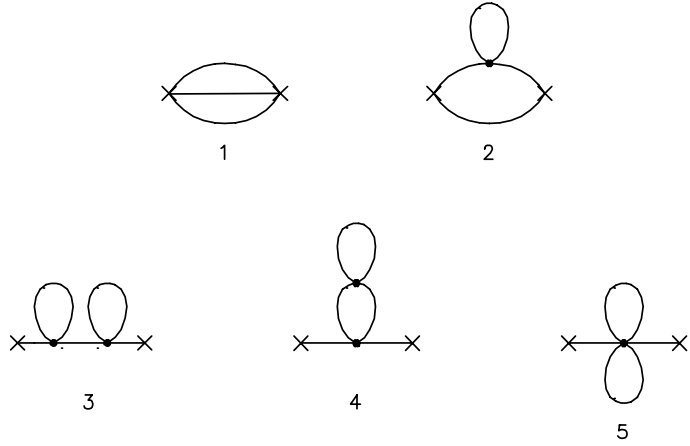}
\vskip 4mm
\end{center}
\caption{Two-loop diagrams contributing to the spin-spin correlation
 function with terms $\propto N^3$}
\label{fig:2loop}
\end{figure}

Out of 12 diagrams only the 5 shown in Fig.~2
have a nonvanishing $O(N^3)$ coefficient. 
Explicitly
\beq
\csi_{3;3}(x,y;L) = \sum_{r=1}^{5} \csi_{3;3}{~}^{(r)}(x,y;L),
\eeq
with
\begin{subeqnarray}
\csi_{3;3}{~}^{(1)} & = &\phantom{-}  \frac{1}{16}
[G(x,x) -G(y,y)]^2\,[G(x,x) + G(y,y)]\;,
\\
\csi_{3;3}{~}^{(2)} & = &  -\frac{1}{8}
 [G(x,x) -G(y,y)]\sum_{\langle i,j \rangle}\;
[G(i,i) -G(j,j)]
\nonumber \\ 
&& \makebox[1.1cm]{ } \times [G(x,i)^2 -G(y,i)^2 - G(x,j)^2
+G(y,j)^2\,]\;, 
\\
\csi_{3;3}{~}^{(3)} & = & \phantom{-} \frac{1}{8}
 \sum_{\langle i,j \rangle\langle k,l \rangle}
\bigg\{ [G(i,i) -G(j,j) ][G(k,k) -G(l,l)]
\nonumber \\ 
&&\makebox[1.1cm]{ }\times G(i,k)\,\Gt(x,y;i)\Gt(x,y;k)
+\;\mbox{3  perms}\;\bigg\}\;,
\\
\csi_{3;3}{~}^{(4)} & = & \phantom{-}  \frac{1}{8}
 \sum_{\langle i,j \rangle\langle k,l \rangle}\bigg\{
[\Gt(x,y;i)^2 -\Gt(x,y;j)^2]
\nonumber\\
&&\makebox[1.1cm]{ }\times [G(k,k) -G(l,l)]\;G(i,k)^2 
+\;\mbox{3  perms}\;\bigg\}\;,  
\\
\csi_{3;3}{~}^{(5)} & = &  -\frac{1}{16}
\sum_{\langle i,j \rangle}\bigg\{\Gt(x,y;i)^2\;
[G(i,i) -G(j,j)][3G(i,i) +G(j,j)] 
\nonumber\\
&& \makebox[3.0cm]{ }
+ (i \leftrightarrow j) \bigg\}\;.
\end{subeqnarray}
Here ``$+\mbox{3 perms}$" stands for three similar terms with summation
indices permuted according to ``$+(i \leftrightarrow j) +
(k \leftrightarrow l) + (i \leftrightarrow j,\;k
\leftrightarrow l)$".

Again we restrict attention to the energy expectation value at the
center of the lattice. First we note that for the analogous
Dirichlet expression one can show
$\lim_{L\to\infty}\cdir_{3;3}(z_0,z_1)=0$.
For the SI functions we again insert the formula, eq.~(\ref{Gxy}),
for $G$ and separate the coefficients of $Z^{-r}$.

Using the results in the last subsection together with the knowledge
that $\lim_{L\to\infty} \sum_{i\ne z_0} \Dt(z_0,z_1;i)^2F(i)=0$ it
is easy to show that the contributions from diagrams 1 and 2 in
Fig.~2 have finite limits as $L\to\infty$. It turns out that
all the remaining diagrams 3--5 in Fig.~2 diverge logarithmically.
Consider first diagrams 3 and 4. Dropping all terms which have finite
$L\to\infty$ limits one obtains the decomposition 
\bea
\!\!\csi_{3;3}{~}^{(3)}(z_0,z_1) \isim s_4 Z^{-4}-32 t_3^2 Z^{-5}\;,
\nonumber\\
\!\!\csi_{3;3}{~}^{(4)}(z_0,z_1) \isim s_2 Z^{-2}+s_3 Z^{-3}+2s_4 Z^{-4}
-32t_3(t_3+t_4Z^2)Z^{-5}\;,
\eea
where $t_4$ is a one-loop sum
\beq
t_4=\frac{1}{4}\sum_{i\ne z_0} D(z_0,i)^2H(i)\;,
\eeq
with
\beq
H(i)=\sum_{j=\langle i\rangle}
\Bigl[ \Dt(z_0,z_1;i)^2-\Dt(z_0,z_1;j)^2 \Bigr]\;.
\eeq
The $s_r$ are more complicated two-loop sums
\bea
s_2\is\frac14\sum_{i\ne z_0}\sum_{k\ne z_0}
H(i)E(k)D(i,k)D(z_0,i)D(z_0,k)\;,
\nonumber\\
s_3\is-\frac{1}{128}\sum_{i\ne z_0}\sum_{k\ne z_0}
E(i)E(k)D(i,k)^2\;,
\nonumber\\
s_4\is \phantom{-}\frac{1}{128}\sum_{i\ne z_0}\sum_{k\ne z_0}
E(i)E(k)D(i,k)D(z_0,i)D(z_0,k)\;.
\eea
We have numerically computed all these sums over a large range
of $L$ (up to $L=210$) and their asymptotic behavior as $L\to\infty$ 
was determined using extrapolation techniques described in the 
previous subsection. In all cases we found results consistent with 
those obtained by substituting the behaviors of the functions at large 
separations:
\bea
&& t_4 \sim \phantom{-}\frac{1}{32}Z\;,
\makebox[2cm]{ }
s_2\sim -\frac{1}{96}Z^{3}\;,
\nonumber\\
&& s_3 \sim -\frac{1}{192}Z^{4}\;,
\makebox[1.6cm]{ }
s_4\sim \phantom{-}\frac{1}{240}Z^{5}\;.
\eea
We do not give all the derivations but illustrate the procedure
again with the interesting case of $s_4$. 
Apart from the approximations (\ref{Dz0i}), (\ref{Ei}) we introduce 
$D(i,k)\sim (1/2\pi) \ln (L/|i-k|)$ so that
\beq
s_4\sim \frac{1}{32}(2\pi)^{-7}\sum_{i\ne z_0}\sum_{k\ne z_0}
{1\over i^2k^2} \ln {L\over|i|}\ln {L\over|k|}\ln {L\over|i-k|}\;.
\eeq
Now by symmetry we can consider $|k|<|i|$ and averaging first over the
angle between $i$ and $k$ we obtain
\bea
s_4\isim \frac{1}{16}(2\pi)^{-7}\sum_{i\ne z_0}
{1\over i^2} \ln {L\over|i|}
\int_{|i|\ge |k|\ge 1}{\rm d}^2k {1\over k^2}
\ln {L\over|k|}\ln {L\over|i-k|}
\nonumber \\
{~}\isim \frac{1}{16}(2\pi)^{-6}\sum_{i\ne z_0}
{1\over i^2} \ln^2 {L\over|i|} \int_1^{|i|} \frac{dr}{r}\ln \frac{L}{r}
\nonumber \\
{~}\is \frac{1}{32}(2\pi)^{-6}\sum_{i\ne z_0}
{1\over i^2} \ln^2 {L\over|i|}\left[ \ln^2 L-\ln^2 \frac{L}{|i|}
\right]
\nonumber \\
{~}\isim \frac{1}{32}(2\pi)^{-5}\Bigl[\frac13-\frac15 \Bigr] \ln^5 L
=\frac{1}{240}(2\pi)^{-5}\ln^5 L\;.
\eea
The corresponding numerical result is shown in Fig.~3.
The fifth logarithmic derivative of $s_4$ multiplied by $(2\pi)^5/5!$
is shown as a function of $1/L$.
\begin{figure}[hbt]
\begin{center}
\leavevmode
\epsfxsize=100mm
\epsfbox{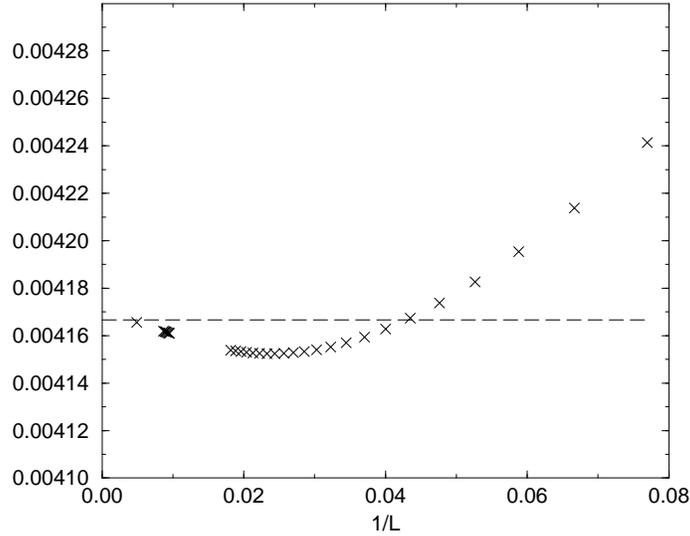}
\end{center}
\caption{The 5-th logarithmic derivative of $s_4$ multiplied by
$(2\pi)^5/5!$ plotted versus $1/L$.}
\label{fig:s4_ld5}
\end{figure}
The definition of the lattice logarithmic derivative used is 
$$
L\frac{\partial}{\partial L}f(L)=
\frac{L}{2}\Big[f(L+1)-f(L-1)\Big]\;.
$$ 
The data are calculated for even values of $L$ in the ranges 
$8-60$, $100-120$, $200-210$ using extended precision arithmetic. 
The dotted line is the analytic result $1/240$.

The coefficient
$\csi_{3;3}{~}^{(5)}(z_0,z_1)$ involves new sums but can be
treated similarly. Summarizing we find
\bea
\csi_{3;3}{~}^{(3)}(z_0,z_1) \isim  \phantom{-}\frac{1}{1440}Z\;,
\nonumber \\
\csi_{3;3}{~}^{(4)}(z_0,z_1) \isim -\frac{1}{2880}Z\;,
\nonumber \\
\csi_{3;3}{~}^{(5)}(z_0,z_1) \isim  \phantom{-}\frac{1}{480}Z\;.
\eea
Thus the divergent contributions do not cancel and one ends up
with an overall logarithmic divergence $\csi_{3;3}(z_0,z_1)\sim 
(7/2880)Z$ as $L\to\infty$.

This shows that one cannot interchange the limits $\beta\to\infty$ 
and $L\to\infty$ for the SI BC because the perturbative 
coefficients diverge at two-loop order. In particular, the SI BC's
do not belong to the class of ``healthy'' ones alluded to in {\bf Q2}
and {\bf Q3} cannot even be addressed. Initially however, the 
SI example was probably thought to discredit PT on internal 
grounds, stemming from the apparent ambiguities in the PT 
answers for the infinite lattice quantities\cite{PSII}. As such, 
we think, it is no longer valid. Of course one could still argue 
that the divergent SI answer might even hit a 
point in signaling that the quantities in question do not have an 
asymptotic expansion in $\beta^{-1}$. However this possibility is 
not an argument by itself and has to be supported by external means,
see, e.g., \cite{PSI}. In the next section we shall leave this example 
and try to develop criteria of what healthy BC's should look like and
under what conditions an affirmative answer to {\bf Q3} should be 
expected.


\section{Addressing the questions {\bf Q2} and {\bf Q3}}
\setcounter{equation}{0}

We start by comparing the correlator between two spins for different
boundary conditions. The discussion will be based on the validity
of the following conjecture: 
For a spin system with ferromagnetic couplings the correlation function 
$C(x,y)=\langle {\bf S}_x \cdot {\bf S}_y \rangle$ increases if any
of the ferromagnetic couplings is increased. 
This conjecture is physically rather intuitive, nevertheless it has not
been proven yet \cite{Sokal}. 
Consider a spin system with a given ferromagnetic interaction on:
\begin{enumerate}
\item a finite lattice $\Lambda$ with free boundary conditions,
\item the same lattice with Dirichlet BC, and
\item an infinite lattice. 
\end{enumerate}
Validity of the above conjecture implies the following inequalities 
between the corresponding correlators:
\beq
C^{\rm free}(x,y;\beta,\Lambda) \le C^{\infty}(x,y;\beta)
\le C^{\rm Dir}(x,y;\beta,\Lambda)\;,
\label{c1}
\eeq
for any $x,y\in\Lambda$. 
More generally, the conjecture also implies that on any lattice
$\Lambda^*$ containing $\Lambda\cup\partial\Lambda$
the correlator $C^*(x,y;\beta,\Lambda^*)$ satisfies 
\beq
C^{\rm free}(x,y;\beta,\Lambda) \le C^*(x,y;\beta,\Lambda^*)
\le C^{\rm Dir}(x,y;\beta,\Lambda)\;.
\label{c1p}
\eeq
From here it follows that $C^{\rm free}(x,y;\beta,\Lambda^*)$
increases while $C^{\rm Dir}(x,y;\beta,\Lambda^*)$ decreases with
increasing size $|\Lambda^*|$. Further the systems with periodic BC
on $\Lambda^*$ will satisfy eq.~(\ref{c1p}).
Note that the SI BC with $z_0\in\Lambda$ (where $z_0$ is the 
site where the extra spin is fixed) does not belong to the class
considered in eq.~(\ref{c1p}), the considerations apply, however,
when $z_0$ is outside $\Lambda$.

In the following we shall argue that if the correlation functions
of the systems 1 and 2 have well-behaved perturbative expansions 
-- in a sense to be specified below -- then
\begin{itemize}
\item[(i)] the infinite system 3 has an asymptotic expansion with
coefficients given by the formal $L\to\infty$ limit of the perturbative
coefficients of systems 1 and 2, and
\item[(ii)] all systems considered under eq.~(\ref{c1p}) have
similarly well-behaved perturbative expansions.
\end{itemize}

For this purpose we have to make some natural and verifiable 
assumptions on the perturbative series with free and Dirichlet BC's.
To simplify the notation we shall return to the square lattice 
of section 2 and write $C^{\alpha}(\beta,L)$ for 
$C^{\alpha}(x,y;\beta,\Lambda)$. The distance of the fixed 
sites $x,y$ from the boundary is taken to be O$(L)$, so that   
$C^{\alpha}(\beta,L)$ qualifies as a short-distance observable, as 
defined in the Introduction. 

Let us elucidate on what we mean by a ``well-behaved'' perturbative
expansion.
For fixed $L$ and given boundary conditions $\alpha=$ free or
Dirichlet, one can calculate $C^{\alpha}$ in PT as a power series 
in $1/\beta$ to some arbitrary order $k$: 
\beq
C^{\alpha}(\beta,L) = 1 - \sum_{r=1}^{k} c^{\alpha}_r(L) \beta^{-r}
+ R^{\alpha}_k(\beta,L)\;.
\label{c2}
\eeq
There is perhaps no doubt that for fixed $L$ these are asymptotic 
series, i.e.
\beq
R^{\alpha}_k(\beta,L)={\rm o}(\beta^{-k})\;. 
\label{c3}
\eeq
Consider now the formal $L\to\infty$ limit of the 
coefficients $c^{\alpha}_r(L)$. On the grounds of (the lattice 
counterpart of) David's analysis \cite{DavidI} one expects that for 
the case of the free, Dirichlet (and periodic) BC's there are no 
infrared singularities and that the convergence to the $L\to\infty$ 
limit is of O($1/L$) (up to logarithms, i.e. O($1/L\cdot \ln^k L$)). 
We shall assume that the perturbative coefficients obtained actually 
coincide in the infinite volume limit:
\beq
c^{\alpha}_r(L)={\bar c}_r+{\rm O}\left( {1\over L}\right), 
~~~\alpha={\rm free, ~Dirichlet},
\label{c4}
\eeq
provided the distance of the fixed sites $x,y$ from the boundary 
is O($L$). One has good reason to trust 
(\ref{c4}). The free propagator for these and other conventional 
BC has a size dependence of  O($1/L^2$) in contrast to the SI case 
with an O($1/\ln L$) dependence. Thus conventional boundary conditions 
should not produce such peculiar behavior as observed in the one-loop 
order for the  SI BC's. Assumption (\ref{c4}) can in principle be 
verified order by order. It could also be relaxed \cite{Sokal} (see 
note added), but we expect eq.~(\ref{c4}) to describe 
the form of the actual finite size corrections. For $r=1,2$ we have 
verified this numerically. Note that $r=2$ is already the source of 
troubles in the SI case.

Clearly, the problem of establishing the claims (i) and (ii) amounts to
controlling the remainders in an ansatz for an asymptotic expansion
of $C^{\infty}(\beta)$ or for the correlators in eq.~(\ref{c1p}). 
Let us first consider claim (i) and define
\beq
R^{\infty}_k(\beta) = C^{\infty}(\beta) - 1 + \sum_{r=1}^k 
\bar{c}_r \beta^{-k}\;.
\eeq
This definition is chosen such that
claim (i) is equivalent to the statement $R_k^{\infty}(\beta) = 
o(\beta^{-k})$, $k \geq 1$. That is $C^{\infty}(\beta)$ has an asymptotic 
expansion in powers of $\beta^{-1}$ and its coefficients 
coincide with $\bar{c}_r$. For later use let us also introduce   
\beq
A^{\alpha}_k(\beta,L) =
\sum_{r=1}^k\left[ c_r^{\alpha}(L) - {\bar c}_r\right]\,
\beta^{-r} \; 
\label{short}
\eeq
and note that one can now rewrite (\ref{c1}) in the form
\beq
A_k^{\rm free}(\beta,L) + R_k^{\rm free}(\beta,L) \leq R_k^{\infty}(\beta)  
\leq A^{\rm Dir}_k(\beta,L) + R_k^{\rm Dir}(\beta,L)\;.
\label{c1a}
\eeq

The naive strategy to establish (i) would be to first try to show that
$|R^{\alpha}_k(\beta,L)|$, $\alpha ={\rm free, Dir}$ are 
bounded by some $L$-independent function of $o(\beta^{-k})$ and then 
conclude from eq.~(\ref{c1a}) that $R^{\infty}_k(\beta) = o(\beta^{-k})$.  
Our main technical observation is that one can reach the same conclusion
using  a bound on $|R^{\alpha}_k(\beta,L)|$ which is much less stringent, 
and which needs to be established only in some part of the $L$-$\beta$ 
plane. In detail, we assume that the bound 
\beq
| R_k^{\alpha}(\beta,L) | \le 
B_k \frac{(\ln L)^{p(k)}}{\beta^{k+1}}\;,
\label{bound}
\eeq
with some finite $B_k$ and $p(k)\ge 0$ has been established in the 
following region of the $L$-$\beta$ plane 
\beq
L \leq L_1(\beta)\,,\;\;\beta > \beta_0\;\;\;{\rm where}\;\;\; 
(\ln L_1(\beta))^{p(k)}/\beta \to 0\;\; {\rm as} \;\;
\beta\to\infty\;\;
\label{lcond1}
\eeq
and $\beta_0$ is some (L-independent) constant. Convenient choices for the 
function $L_1(\beta)$ will be given below. Similarly define another 
region of the $L$-$\beta$ plane by
\beq
L_0(\beta)\leq L\,,\;\;\beta > \beta_0\;\;\;{\rm where}\;\;\; 
\beta^{k-1}/L_0(\beta) \to 0\;.
\label{lcond2}
\eeq
Choosing now any unbounded path $L(\beta)$ in the intersection
of both regions (\ref{lcond1}) and (\ref{lcond2}) the 
desired conclusion can be reached immediately: One has 
$A_k^{\alpha}(\beta,L(\beta))=o(\beta^{-k})$ by eq.~(\ref{c4}) and the 
condition on $L_0(\beta)$. Similarly from eq.~(\ref{bound}) and the 
condition on  $L_1(\beta)$ it follows that 
$R_k^{\alpha}(\beta,L(\beta))=o(\beta^{-k})$. 
Then the inequalities (\ref{c1a}) imply
\beq
R_k^{\infty}(\beta)=o\left( \beta^{-k}\right) \;,
\label{asym}
\eeq
which is basically the required result. Strictly speaking, one should 
for a fixed $k$ refer to the numbers $\bar{c}_r,\;r \leq k$ as the 
(unique) candidates for the coefficients in an asymptotic expansion 
of $C^{\infty}(\beta)$ in $\beta^{-1}$. 
Only after eq.~(\ref{bound}) and hence eq.~(\ref{asym}) has been shown 
for all $k$, 
the existence of the asymptotic expansion follows and the numbers  
$\bar{c}_r$ can properly be referred to as the coefficents 
of this expansion, so that claim (i) follows. It is also easy to 
see that starting with eq.~(\ref{c1p}) the same strategy applies and 
eq.~(\ref{bound}) also implies claim (ii). 

Let us add a number of comments. First one should check that
choices for $L_0(\beta)$ and $L_1(\beta)$ exist such that the 
intersection of the regions (\ref{lcond1}) and (\ref{lcond2})
is nonempty for each sufficiently large $\beta$.
A simple example is $L_0(\beta) = L_1(\beta) =\beta^k$,
which is moreover the choice with about the minimal growth that 
allows one to draw conclusions about the first $k$ coefficients of the
expansion. Of course any choice for $L_0(\beta)$ and $L_1(\beta)$ 
growing faster than $\beta^k$ (and such that the intersection of the 
regions (\ref{lcond1}) and (\ref{lcond2}) is nonempty) is also
sufficient. For example a region containing the path  
\beq
L(\beta) = \exp \left( \ln^2 \beta \right) 
\label{eln2}
\eeq 
would be convenient, because it allows one to cover all $k$ 
simultaneously in the above argument. However one should keep in mind 
that the faster the upper boundary $L_1(\beta)$ grows, the more difficult 
it may be to establish the bound (\ref{bound}) within the region
(\ref{lcond1}). 

A second comment concerns the relation to David's theorem.
According to (the lattice counterpart) of this result one expects that 
the infrared divergent terms cancel in the perturbative coefficients.
However, to directly prove such cancellations in the
remainder $R_k^{\alpha}(\beta,L)$ (e.g., $p(k)=0$ in eq.~(\ref{bound}))
is probably a more difficult task. The point of using the correlation 
inequalities is twofold: First, the claim (i) can be established without
assuming the IR finiteness of the remainders for fixed $\beta$, 
as described before. And second, once claim (i) is established, the IR 
finiteness of the remainders actually follows from the correlation 
inequality (\ref{c1p}). To see the latter let $L(\beta)$ be an unbounded 
path in the intersection of the regions (\ref{lcond1}), (\ref{lcond2})
and let $(\beta,L')$ be any point in the region 
$\{(\beta,L'')\,|\, \beta >\beta_0,\,L'' >L(\beta)\}$. 
Then by (\ref{c1p}) 
\begin{eqnarray}
A_k^{\rm free}(\beta,L(\beta)) + R_k^{\rm free}(\beta,L(\beta)) \leq 
A_k^{\rm free}(\beta,L') + R_k^{\rm free}(\beta,L') \leq 
R_k^{\infty}(\beta) \nonumber\\
\leq
A^{\rm Dir}_k(\beta,L') + R_k^{\rm Dir}(\beta,L') \leq
A_k^{\rm Dir}(\beta,L(\beta)) + R_k^{\rm Dir}(\beta,L(\beta))\;.
\end{eqnarray}
Since, given eq.~(\ref{bound}), we know the behavior $o(\beta^{-k})$ for 
the terms with $L(\beta)$ and by eq.~(\ref{c4}) also for 
$A_k^{\alpha}(\beta,L')$, it follows that
$R_k^{\alpha}(\beta,L')=o(\beta^{-k})$ 
is valid ``uniformly'' in the region $L' \geq L(\beta)$.
(In other words, the statement is true for arbitrary path 
$L'(\beta)>L(\beta)$ with a bound on $R_k^{\alpha}(\beta,L'(\beta))$
which is independent on how fast $L'(\beta)$ grows.
Note that this is not uniformity in the usual sense where one would 
require a region $L' \geq L'_0$ with $L'_0$ independent of $\beta$.)
In a sense this supplements (the expected lattice 
counterpart of) David's argument on the IR finiteness of the   
PT coefficients. 

It remains to establish eq.~(\ref{bound}). Since we have not managed 
to prove this, the proof of (i) and (ii) remains incomplete.
However, we find the strategy outlined promising and for 
the rest of this section we shall give a plausibility
argument that eq.~(\ref{bound}) indeed holds. To explain the 
argument let us follow the expectation of Patrascioiu and Seiler 
that PT breaks down for large lattice sizes $L$ since the spins at 
large relative distances become strongly decorrelated. Consequently, 
one can question the validity of the bound (\ref{bound}) in the whole 
$L$-$\beta$ plane, in particular for $\beta$ fixed, $L\to\infty$.
Recall, however, that in the strategy we used to establish claim
(i) exclusively lattice sizes of type (\ref{eln2}) enter which grow 
with $\beta$ only in 
lattice units -- the physical size of these lattices $L(\beta)/\xi(\beta)$
decreases as $\beta\to\infty$. (Here $\xi(\beta)$ is the correlation
length growing exponentially in the standard scenario.)
Consequently, in our stratgey it is sufficient to establish
the bound (\ref{bound}) in the region (\ref{lcond1}) of the $L$-$\beta$
plane. As it is shown below, for such mildly growing volumes the spins
become increasingly ordered with growing $\beta$ -- hence the above
objection does not apply.
Roughly speaking, the argument estimates the probability of having 
fluctuations $\vec{\pi}_x^2 \ge \delta^2$ at least at one of 
the sites and finds that such fluctuations are exponentially 
suppressed by a factor $\sim \exp(-\pi\delta^2\beta/\ln L)$.

We begin by considering the effect of constraining 
the fluctuations in an auxiliary system.
Let us calculate the probability that in a massless free
theory the fluctuation exceeds some given threshold $\delta$.
We shall use again the $\Lambda \cup \partial\Lambda$ square lattice 
of section 2 and the action
\beq
{\cal A}_0(\phi)={1\over 2}\sum_{x,\mu} \left(
\nabla_{\mu}\phi_x\right)^2.
\label{c7}
\eeq
In the case of the Dirichlet BC the field at the boundary is fixed
to zero, while for the free BC we use the remaining global
translational symmetry to fix, say $\phi_0 \equiv \phi(x=0)=0$.
(For this BC the site $x=0$ is assumed to be left out in 
the sums and products appearing in the expressions below.)
Consider now the constrained model described by the partition function
\beq
Z(\beta,\delta)=\int_{-\delta}^{\delta}
\prod_{x\in \Lambda} d\phi_x e^{-\beta {\cal A}_0(\phi)}.
\eeq
Here the field values are restricted to $|\phi_x| \le \delta$.
Define the correction term $\bar{R}(\beta,\delta)$ as
\beq
Z(\beta,\delta)=Z(\beta)\left[ 1- \bar{R}(\beta,\delta) \right] \, ,
\label{c8}
\eeq
where $Z(\beta)=Z(\beta,\infty)$ is the corresponding partition
function for the unrestricted case. The term $\bar{R}(\beta,\delta)$ 
can be interpreted as the probability to have $|\phi_x| > \delta$,
for at least one of the sites $x$.
It satisfies the inequality:
\begin{eqnarray*}
&&
 0 \le \bar{R}(\beta,\delta) \le \frac{1}{Z(\beta)}
\sum_{x\in \Lambda} \int_{|\phi_x|\ge \delta} d\phi_x
\int_{-\infty}^{\infty} \prod_{y\ne x} d\phi_y
e^{-\beta {\cal A}_0(\phi)}
\nonumber \\
&&
= \frac{1}{Z(\beta)}
\sum_{x\in \Lambda} \frac{1}{2\pi i}\int_{-\infty}^{\infty}d\alpha
\left( \frac{ e^{i\alpha\delta}}{\alpha +i0} -
 \frac{ e^{-i\alpha\delta}}{\alpha -i0} \right)
\int_{-\infty}^{\infty} \prod_y d\phi_y
e^{i\alpha \phi_x - \beta {\cal A}_0(\phi)}
\nonumber \\
&&
= \sum_{x\in \Lambda} 
\frac{1}{2\pi i}\int_{-\infty}^{\infty}d\alpha
\left( \frac{ e^{i\alpha\delta}}{\alpha +i0} -
 \frac{ e^{-i\alpha\delta}}{\alpha -i0} \right)
{\rm exp}\left( -\frac{D_{xx}\alpha^2}{2\beta}\right)
\nonumber \\
&&
= \sum_{x\in \Lambda} {\rm exp}
\left( -\frac{\beta\delta^2}{2 D_{xx}}\right)
F\left(\delta\sqrt{ \frac{\beta}{D_{xx}} } \right) \, .
\end{eqnarray*}
Here $D_{xx}=\langle \phi_x^2\rangle$. The function $F$ is given by
$$
F(\delta)=\frac{1}{\pi}\int_{-\infty}^{\infty}du\,e^{-u^2/2}
\frac{\delta}{u^2+\delta^2} \, ,
$$
and has the properties $F(0)=1$, $F(\delta)\le 1$, and $F(\infty)=0$. 
Finally, one has
\beq
0 \le \bar{R}(\beta,\delta) \le |\Lambda| 
{\rm exp}\left( -\frac{\beta\delta^2}{2D_{{\rm max}}} \right) \, ,
\label{Rbound1}
\eeq
where $D_{{\rm max}}=\max_x D_{xx}$. In two dimensions
$D_{{\rm max}}\sim (2\pi)^{-1} \ln L$ hence
\beq
0 \le \bar{R}(\beta,\delta) \le L^2\, 
{\rm exp}\left( -\frac{\pi\beta\delta^2}{\ln L} \right)\,.
\label{Rbound2}
\eeq
In fact, $\delta^2/2 D_{xx}$ is the minimal value of the action under
the condition that $\phi_x=\delta$.
To show this consider the configuration
\beq
\phi_y= \frac{D_{xy}}{D_{xx}}\, \delta  \quad 
{\rm for\,\, any}\,\,y \, .
\label{SIsol}
\eeq
Obviously, $\phi_x=\delta$ and it satisfies the lattice equations of
motion, $\Delta\phi_y=0$ for any $y\ne x$.
The corresponding action value is given by
\beq
 -\frac{1}{2} \sum_y \phi_y (\Delta \phi)_y
= \frac{\delta^2}{2 D_{xx}} \, ,
\label{SIaction}
\eeq
as stated above.
It is also easy to show that the $n$-point functions in the constrained
model will differ from those in the unconstrained one also by
exponentially small terms of the form (\ref{Rbound2}).

We now return to the O($n$) $\sigma$ model. Again, for the case of 
the free BC's we use the global O($n$) symmetry to fix 
${\bf S}_0 =(\vec{0},1)$. Using the parametrization
${\bf S}_x=(\vec{\pi}_x,\pm\sqrt{1- \vec{\pi}_x^2})$ 
it is easy to verify the following inequality:\footnote{Note 
that for $S^n>0$ the parametrization
${\bf S}=(\vec{\phi},1)/\sqrt{1+\vec{\phi}^2}$ gives an upper
bound, $1-{\bf S}_x \cdot {\bf S}_y \le 
\frac{1}{2} (\vec{\phi}_x - \vec{\phi}_y)^2$, which also might be 
useful.}
\beq
1-{\bf S}_x \cdot {\bf S}_y \ge 
\frac{1}{2} (\vec{\pi}_x - \vec{\pi}_y)^2 \, .
\label{ineq1}
\eeq
As a consequence of eq.~(\ref{ineq1}), the standard action of 
the $\sigma$ model is bounded from below by the corresponding 
free field action, 
${\cal A}({\bf S}) \ge {\cal A}_0(\vec{\pi})$ and
\beq
{\cal A}({\bf S}) \ge \frac{\delta^2}{2 D_{xx}} 
\quad {\rm for}\quad |\vec{\pi}_x | \ge \delta \, .
\label{Amin}
\eeq
From here it follows that the probability to have 
$|\vec{\pi}_x| > \delta$ at least at one site $x$ is suppressed by 
the exponential factor given in eqs.~(\ref{Rbound1}) and (\ref{Rbound2}).
Once again, the constraint influences the correlation functions 
only by an exponentially small amount as $\beta\to\infty$.
As anticipated, increasing the lattice size not too fast, 
for example, as in eq.~(\ref{eln2}), the correlation functions
of the constrained and unconstrained systems will differ by correction
terms which decrease faster than any inverse power of $\beta$. 
Further the bound (\ref{Rbound2}) allows one to choose a constraint 
$\delta$ which decreases with increasing $\beta$ not too fast, 
say as $\delta(\beta)^2 \sim \ln^3 \beta \ln L/\beta$.
In this case the bound still vanishes faster than any inverse power 
of $\beta$, while the quantity $\beta \vec{\pi}^4$, the leading 
perturbative term in $\beta{\cal A}({\bf S})$, goes to zero as 
$\beta\to\infty$.

By comparing the constrained and unconstrained systems one concludes
that the fluctuations in the unconstrained system are essentially
bounded as $\vec{\pi}_x^2 \le {\rm const~} \ln L/ \beta$.
In other words a system of mildly growing size $L(\beta)$ 
(e.g., as in eq.~(\ref{eln2})) becomes increasingly ordered as 
$\beta\to\infty$. As shown before, within our strategy it is sufficient 
to prove the existence of the bound (\ref{bound}) for such mildly growing
$L(\beta)$. By the above reasoning this should be a simpler task, 
but still remains to be done.

\section{Discussion}
\setcounter{equation}{0}

Let us first discuss the physical picture behind the arguments 
of the previous section in more detail.
In \cite{PSII} the so-called superinstanton configurations 
are introduced. In analogy with the free field case one considers 
the configuration with minimal action under the condition that 
the spin at the middle (at $x=z_0$) is rotated by an arbitrary angle 
relative to the spins at the boundary (fixed by the Dirichlet BC). 
The action value of these configurations is O($1/\ln L$) in two 
dimensions, so that they can be viewed as saturating the bound 
(\ref{Amin}). They play a crucial role in disordering the system 
in the infinite volume. As the authors note, the fact that the energy 
of superinstantons goes to zero as $L\to\infty$ implies that, in an 
infinite volume, they are present at arbitrarily large $\beta$ and 
disorder the spins forbidding a spontaneous magnetization in two dimensions.
With the assumptions made we have shown, however, that for establishing 
the correctness of the perturbative expansion it is enough to consider
a mildly growing size with $L(\beta)$ given, e.g., by eq.~(\ref{eln2}).
Under such circumstances the superinstantons with
$|\vec{\pi}_x| > \delta$ are exponentially suppressed 
and the error made by restricting the integration region to
$|\vec{\pi}_x| \le \delta$ is exponentially small. 
(After the standard PT steps -- rescaling $\vec{\pi}_x$ and expanding
the Boltzmann factor with only the quadratic part kept in the exponent
-- the dependence on $\delta$ is again exponentially suppressed,
as expected.) 
For the correlation function $\langle S_x\cdot S_y\rangle$ only 
superinstantons
with size smaller than $|x-y|$ contribute significantly, larger ones
will rotate the two spins simultaneously. This contribution is responsible 
for the leading order bare PT result 
$$
\langle ( {\bf S}_x - {\bf S}_0 )^2 \rangle \sim
\frac{1}{\beta} \frac{N}{2\pi} \ln |x| \;.
\label{SPT}
$$
This also explains why the O(n) invariant quantities are infrared 
finite in PT, as opposed to the noninvariant ones which diverge as 
some power of $\ln L$. However the form of the finite size corrections
is incorrectly given by PT, eq.~(\ref{c4}).
Indeed, due to the nonperturbatively generated mass the corrections 
should be exponentially small for $L\gg \xi(\beta)$. 

In $d=1$ dimensions the status of PT is quite different. 
In one dimension $D_{\rm max} \propto L$ and the correction term is of 
the form $\exp(-c\beta\delta^2/L)$. Consequently, the O($1/L$) corrections
in eq.~(\ref{c4}) are not negligible compared to higher order terms 
in $1/\beta$ even for such large sizes when the perturbative expansion 
breaks down, i.e., $L(\beta)\propto \beta$. This is in accordance with 
the observation by Hasenfratz \cite{Hasenfratz} that the limits 
$\beta\to\infty$ and $L\to\infty$ in $d=1$ are not interchangeable.
As pointed out in \cite{Brezin}, starting from order $\beta^{-3}$ 
the coefficients are infrared divergent.

The case of the  SI BC in $d=2$ is very similar to the $d=1$ case
with the usual BC. Indeed, already the tree level result
has a finite size correction O($1/\ln L$) which is comparable 
to higher order terms even for exponentially large $L(\beta)$
where PT breaks down.
The correlation inequalities for this case read
\beq
 C^{\infty}(x,y;\beta)
\le C^{\rm Dir}(x,y;\beta,L) \le C^{\rm SI}(x,y;\beta,L).
\label{c15}
\eeq
The first few terms of the perturbative expansion for the nearest
neighbors are given by \cite{PSII}
\bea
C^{\rm SI}(z_0,z_1;\beta,L) \is 1 -
{N\over \beta}\left[\frac{1}{4} - \frac{\pi}{16\ln L} + 
{\rm O}\left( {1\over \ln^2 L} \right) \right] \nonumber\\
&& \phantom{1} - {N\over \beta^2}\left[\frac{N}{48} + \frac{1}{96} +
{\rm O}\left( {1\over \ln L} \right) \right] + \ldots
\label{c16}
\eea
while the corresponding piece of $C^{\infty}$ is
\beq
C^{\infty}= 1-{N\over 4\beta}-{N\over 32\beta^2} +\ldots
\label{c17}
\eeq
Clearly, eq.~(\ref{c16}) is inconsistent with the correlation inequality
(\ref{c15}) for $N>1$ (i.e., $n=N+1>2$)
when the formal $L\to\infty$ limit is taken.
In the regime $\ln L(\beta) \le \ln \xi(\beta) \propto\beta$, however, 
the $1/\ln L$ correction term in the tree level 
contribution restores the inequality.

What is the origin of the anomalously large finite size correction for
the SI BC? 
Consider a general SI BC  where the spin ${\bf S}_{z_0}$ is fixed 
to a direction which is not necessarily parallel to the spins on 
the boundary $\partial\Lambda$.
Denoting the corresponding expectation value by
$\langle {\cal O} \rangle_{{\bf S}_{z_0}}$, the following
relation holds 
\beq
\langle {\cal O} \rangle_{\rm Dir} =
\frac{1}{Z(\beta)} \int d{\bf S}_{z_0} 
\langle {\cal O} \rangle_{{\bf S}_{z_0}} 
e^{-\beta F({\bf S}_{z_0})} \;,
\eeq
where $F({\bf S}_{z_0})$ is the free energy of a superinstanton
and $Z(\beta)$ is the partition function for Dirichlet BC%
\footnote{A curious point in the terminology is
that the SI BC with ${\bf S}_{z_0}=(\vec{0},1)$ is the one which
{\em excludes} superinstantons centered at $z_0$. Rather  
the ordinary BC takes care of {\em all} superinstantons.}
In the lowest order in $1/\beta$ and for $|\vec{\pi}_{z_0}| \ll 1$
the free energy is up to a (for our purposes irrelevant) additive 
constant given by (cf. eq.~(\ref{SIaction}))
\beq
F({\bf S}_{z_0}) = \frac{1}{2 D(z_0,z_0)}\vec{\pi}_{z_0}^2 \,.
\label{F_SI}
\eeq
The SI solution gives (cf. eq.~(\ref{SIsol}))
\beq
\vec{\pi}_{z_0}-\vec{\pi}_{z_1} \approx \frac{1}{4 D(z_0,z_0)}
\vec{\pi}_{z_0} \,.
\eeq
Then the integration over the SI solutions (still without 
the contribution from the fluctuations) introduces a nontrivial 
expectation value for ${\bf S}_{z_0} \cdot {\bf S}_{z_1}$ given by
\beq
 1 - {N \pi \over 16 \beta \ln L} +
{\rm O}\left( {1\over \beta \ln^2 L} \right) \,.
\label{c18}
\eeq
The analogous correction term in eq.~(\ref{c16}) --- which
is a result of the fluctuations for the case when ${\bf S}_{z_0}$
is fixed --- compensates the large finite size correction in 
eq.~(\ref{c18}), as it should, since no such term appears for the
Dirichlet BC, eq.~(\ref{c1Dir}). 
As eq.~(\ref{F_SI}) shows, the natural expansion parameter
by integrating over the SI directions is $\ln L/\beta$, and
it is easy to see that in generic O$(N+1)$ models the higher order 
contributions in eq.~(\ref{c18}) will be of the form 
$1/\ln^2 L \cdot (\ln L/\beta)^k$.
As a consequence, a logarithmic divergence in order $1/\beta^3$ 
is expected also in the perturbative result with the SI BC,
in agreement with the explicit computation of section 2.
Since only the coefficient of the leading power in $N$ has been 
computed one cannot exclude that for a specific $N$ the 
divergence is cancelled against one from a subleading power in $N$.
In particular it would be interesting to see what happens in the 
O$(2)$ model.

So far we mainly considered bare lattice PT in volumes
whose physical size goes to zero in the continuum limit,
$L(\beta)/\xi(\beta)\to 0$ as $\beta\to\infty$. 
However, this was sufficient, together with the correlation 
inequalities, to argue that the standard bare PT provides 
the correct asymptotic expansion for the system in an infinite 
lattice. The physically more relevant question is whether the 
renormalized PT provides an asymptotic expansion for the correlation 
function at short physical distances, $x_{\rm phys}= \epsilon a \xi$,
$0 < \epsilon \ll 1$. A positive answer to this question is not
an automatic consequence of the proposed validity of 
the bare lattice PT.

Our considerations did not make use of the ``integrability'' of the 
${\rm O}(n)$ models. Assuming asymptotic freedom one can establish the 
existence of a Yangian algebra of non-local conserved charges and the 
absence of particle production\cite{Lu, NLC}. The ${\rm O}(n)$ symmetry 
then basically determines the $S$-matrix amplitudes \cite{Zamol}. This 
bootstrap $S$-matrix has been tested (at low energies) in lattice
studies \cite{LuWo} and 
used as an input for the thermodynamic Bethe ansatz to compute 
the exact $m/\Lambda$ ratio \cite{HaNi}. 
The results are also consistent with the $1/n$ expansions \cite{Rossi}.
Finally one can use the bootstrap $S$-matrix as an input for the form 
factor approach \cite{KaWe, Smir} which provides an alternative 
nonperturbative definition of the theory. 
For the ${\rm O}(3)$ model the results in \cite{BaNi} strongly 
indicate that the model thus constructed coincides with the continuum 
limit of the lattice theory at least at low energies. At intermediate 
energies the results coincide with renormalized PT and at high 
energies they are consistent with asymptotic freedom.
Two non-perturbative constants determined exactly in this approach 
\cite{BaNiII} are again consistent with Monte Carlo data 
\cite{Alles,Pisa,BaNiII}.
One cannot help feeling that the most natural way to reconcile these 
facts is the conventional wisdom. That is the continuum limit of the 
lattice theory coincides with the model described by the bootstrap 
approach, which is in turn correctly described by an asymptotically 
free PT at high energies. 

{\it Acknowledgements}
Firstly we would like to express our particular thanks to Alan Sokal
for numerous discussions and helpful comments, and for sending us
his proof of a related theorem (see below). We would also like
to thank Peter Hasenfratz, Adrian Patrascioiu and Erhard Seiler
for useful discussions. M.N. acknowledges support 
by the Reimar L\"ust fellowship of the Max-Planck-Society. 
A first version of this paper stimulated the response
hep-lat/9702008 by A. Patrascioiu and E. Seiler, which in 
turn motivated us to improve on our exposition. 


{\it Note added}
A concise alternative to our first exposition is also contained 
in a recent note by Sokal \cite{Sokal}. In particular he observes that 
the assumption (\ref{c4}) on the finite size
dependence of the coefficients can be relaxed as follows:
Assume that there exist coefficients $\{ \bar{c}_r \}_{r=1}^k$
and powers $\epsilon(r)>0$ such that
$$
| c_r^{\alpha}(L)-\bar{c}_r | \le O\left( L^{-\epsilon(r)} \right)\;,
\mbox{~~~$\alpha$=free, Dir}
$$
for $r=1,\ldots,k$. Further replace the condition on $L_0(\beta)$ in
eq.~(\ref{lcond1}) by $\beta^{k-r}L_0(\beta)^{-\epsilon(r)}\to 0$
for $r=1,\dots,k$.
He assumes that the bound (\ref{bound}) has been established in
the whole $L$-$\beta$ plane, but this could also be weakened along
the lines described in section 3. Thus assuming that the bound 
(\ref{bound}) has been established for at least one unbounded path 
contained in the new region (\ref{lcond1}) he shows that 
$C^{\infty}(\beta)$ has an asymptotic expansion with the coefficients 
$\bar{c}_r$. The proof remains essentially the same: One takes an $L(\beta)$  
for which both $A_k^{\alpha}$ and $R_k^{\alpha}$ are 
$o(\beta^{-k})$ under the assumptions stated and concludes 
$R_k^{\infty}(\beta) = o(\beta^{-k})$ from the correlation inequality
(\ref{c1}).

\vspace{1cm}


\eject

\end{document}